\documentclass[twocolumn,prd,superscriptaddress,showpacs,amsmath,amssymb]{revtex4-1}

\usepackage{graphicx}
\usepackage{dcolumn}
\usepackage{bm}
\usepackage{multirow}

\begin{document}

\title{Applications of an $^{88}$Y/Be photo-neutron calibration source \\ to Dark Matter and Neutrino Experiments}

\def\UC{Kavli Institute for Cosmological Physics and Enrico Fermi Institute, University of Chicago, Chicago, IL 60637, USA}

\author{J.I.~Collar} \email{Electronic address: collar@uchicago.edu}\affiliation{\UC}
\date{\today}

\begin{abstract}
The low-energy monochromatic neutron emission from an $^{88}$Y/Be source can be exploited to  mimic the few keVnr nuclear recoils expected from low-mass Weakly Interacting Massive Particles (WIMPs) and coherent scattering of neutrinos off nuclei. Using this source, a $\lesssim$10\% quenching factor is measured for sodium recoils below 24 keVnr in NaI[Tl]. This is considerably smaller than the 30\% typically adopted in the interpretation of the DAMA/LIBRA dark matter experiment, resulting in a marked increase of its tension with other searches, under the standard set of phenomenological assumptions. The method is illustrated for other target materials (superheated and noble liquids).
\end{abstract}



\maketitle

Several recently observed anomalies in  direct searches for dark matter \cite{DAMA,cogent,cresst} can be interpreted within the context of low-mass (few GeV/c$^{2}$) Weakly Interacting Massive Particles (WIMPs). These hypothetical particles are expected to induce nuclear recoils \cite{goodman} carrying just a few keV (keVnr) of energy in most targets. A recoiling nucleus produces a reduced amount of ionization or scintillation in detector materials, when compared to an electron recoil of the same energy. Careful study of this "quenching factor" is a necessary first step in the interpretation of dark matter detector data. 

A scarce number of calibration methods exist for this few keVnr nuclear recoil energy regime, and those typically involve the use of accelerator or reactor facilities able to provide monochromatic neutron beams, and of small dedicated detectors. This Letter introduces a convenient new method able to generate such recoils. The source is eminently fieldable, i.e., can be used {\it in situ} at large underground dark matter detectors, in this way incorporating specific instrumental effects into the measured response. Planned devices aiming at detection of coherent neutrino scattering off target nuclei \cite{freedman,cosi} can also benefit from its characteristics. 
 
Photo-neutron sources involve the gamma-induced disintegration of a light nucleus with modest neutron binding energy, such as $^{9}$Be or $^{2}$D \cite{report,knoll}. Only a few relatively short-lived radioisotopes generate gammas above threshold for these reactions (1.67 MeV and 2.23 MeV, respectively). The main advantage of these sources is the mono-energetic nature of the neutrons generated, typically in the few hundreds of keV. However, their neutron yield is several orders of magnitude lower than their gamma counterpart, due to the relatively small cross-section for the photo-disintegration reaction. On first impression, this large unbalance would complicate their use for calibration of WIMP detectors, specially when considering that the ability to discriminate between gamma-induced electron recoils and neutron-induced nuclear recoils is often very limited at the few keVnr expected.

\begin{figure}[!htbp]
\includegraphics[width=0.45\textwidth]{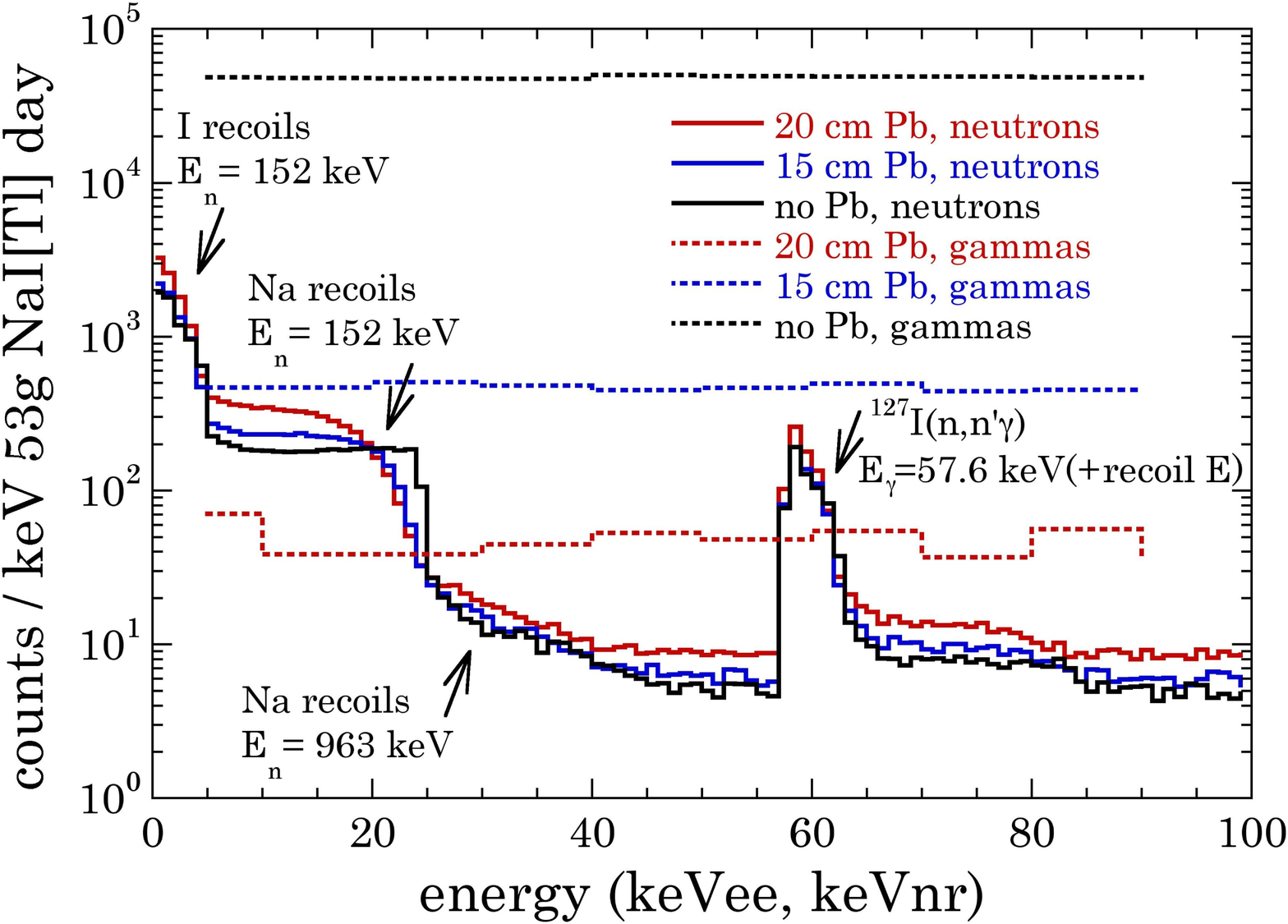}
\caption{\label{fig:sodium} MCNP-PoliMi \protect\cite{polimi} simulated deposited energy by electron recoils (keVee) and nuclear recoils (keVnr) in a small NaI[Tl] scintillator irradiated by $^{88}$Y/Be gammas and neutrons, respectively. The 0.1 mCi source is 21 cm away from the detector, yielding an isotropic 400 neutrons/s and 7.4$\times 10^{6}$ gammas/s. Several intermediate thicknesses of lead between source and detector are considered (see text).}
\end{figure}

A way to bypass this issue is the addition of 15-20 cm of lead shielding around the source. Conveniently, this is a typical thickness already present around many large underground WIMP detectors. Lead combines optimal gamma shielding properties with one of the smallest neutron moderation powers in any material, due to its large atomic mass. Formally defined, this power is the product of neutron lethargy times the macroscopic scattering cross section, $2\times10^{-3}$ for lead and unity for the standard moderator, hydrogen \cite{moderator}. More intuitively, a $\sim$100 keV neutron straggles in lead with a mean free path of $\sim$3 cm between scatters, while losing just a maximum of 2\% of its energy in each interaction, typically less. 

Fig.\ 1 exemplifies this approach for the case of a small NaI[Tl] scintillator in the presence of an $^{88}$Y/Be source. The combination of $^{88}$Y (T$_{1/2}$= 107 d) and a beryllium converter produces a dominant E$_{n}$=152 keV neutron emission with just a $\sim$0.5\% E$_{n}$=963 keV higher-energy neutron branch, from gamma rays at E$_{\gamma}$=1.836 MeV and 2.734 MeV, respectively \cite{knoll}.  The maximum recoil energy imparted to a sodium nucleus by the dominant neutrons is kinematically limited to (4~M~m~E$_{n}$)/(M+m)$^{2}=$ 24 keVnr, with M and m representing the sodium nucleus and neutron rest mass energies, respectively. The lead layer reduces the gamma flux arriving to the detector under test to manageable levels, while neutrons lose only a small fraction of their energy in traversing it.

\begin{figure}[!htbp]
\includegraphics[width=0.45\textwidth]{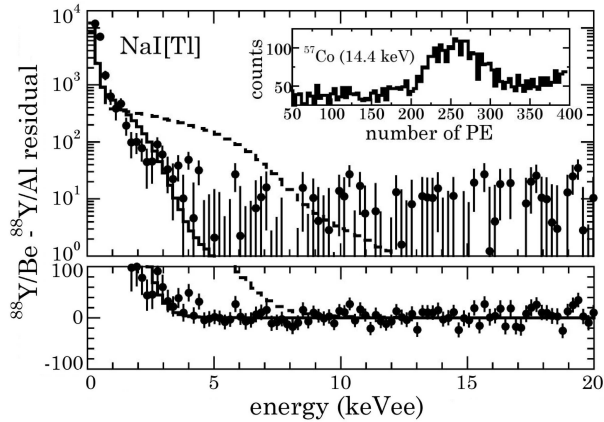}
\caption{\label{fig:measurements} Measured response of NaI[Tl] to low-energy nuclear recoils from an $^{88}$Y/Be source (see text). A solid histogram is the predicted response, allowing no free parameters, obtained by adopting the quenching factors for sodium and iodine recoils recently measured using 2.2 MeV neutron scattering from a D-D generator (a monotonically decreasing $Q_{Na}$ with decreasing recoil energy, and $Q_{I}\!=\!$ 0.04 \cite{DD}). A dashed histogram, in large disagreement with present data, employs quenching factors typically used in the interpretation of DAMA/LIBRA results ($Q_{Na}\!=\!$ 0.3,  $Q_{I}\!=\!$ 0.09 \protect\cite{damaquenching}).}
\end{figure}

Fig.\ 2 demonstrates the experimental application of this concept. A small (1.9 cm diameter, 5.1 cm long) NaI[Tl] scintillator from Proteus/Amcrys was surrounded by 5 cm of lead, except for on one side of a rectangular box enclosure, 20 cm thick. Outside of this thicker lead wall, a small cylindrical "pillbox" (4.8 cm diameter, 2.5 cm tall) of BeO obtained from American Beryllia was used to encapsulate an $^{88}$Y button source, procured from Eckert \& Ziegler Isotope Products.  The activity of this source at the time of the measurement was approximately 50 $\mu$Ci, with an isotropic neutron yield of $\sim$330 n/s, characterized with a $^{3}$He counter surrounded by polyethylene moderator, the response of which was calculated via MCNP simulation. 

The same data acquisition (DAQ) system as in \cite{DD} was employed, set to trigger on single photoelectrons (PE) from a Hamamatsu ultra-bialkali photomultiplier directly coupled to the scintillator. The rapid triggering rate ($\sim$800 Hz in the presence of the source) was well-below the maximum throughput  of this DAQ. An energy scale was established using the 14.4 keV gamma emission from a $^{57}$Co source (Fig.\ 2 inset), resulting in a light yield, integrated over 3 $\mu$s, in good agreement with that found in \cite{DD} via Compton scattering. These gammas are sufficiently penetrating to clear the 0.8 mm scintillator aluminum casing, while also avoiding skin effects from hydrated surface layers in the crystal \cite{gerbier}. 

Four pairs of runs were taken, with the $^{88}$Y source surrounded alternatively by BeO or by a pillbox replica in aluminum metal, for a total of 90 hours in each configuration. Aluminum has gamma cross sections essentially identical to BeO for penetrating photons from this radioisotope, but is inert for purposes of neutron emission. Two cuts were applied to the data during their analysis. First, an energy-dependent dead time was imposed following large energy depositions in the crystal, so as to avoid excess low-energy signals from the ensuing afterglow (phosphorescence). Second, only events with a scintillation decay time $\tau$ compatible with that from nuclear and electron recoils were accepted. This range is 50 ns $<\!\tau\!<$ 400 ns for NaI[Tl] \cite{DD}. Fig.\ 2 displays the cumulative residual energy spectrum obtained by taking the difference between  $^{88}$Y/Be (neutron plus gamma) and  $^{88}$Y/Al (gamma only) runs. The low-energy excess from neutron-induced nuclear recoils is readily visible in this residual, and present for all pairs of runs. 

\begin{figure}[htbp]
\includegraphics[width=0.45\textwidth]{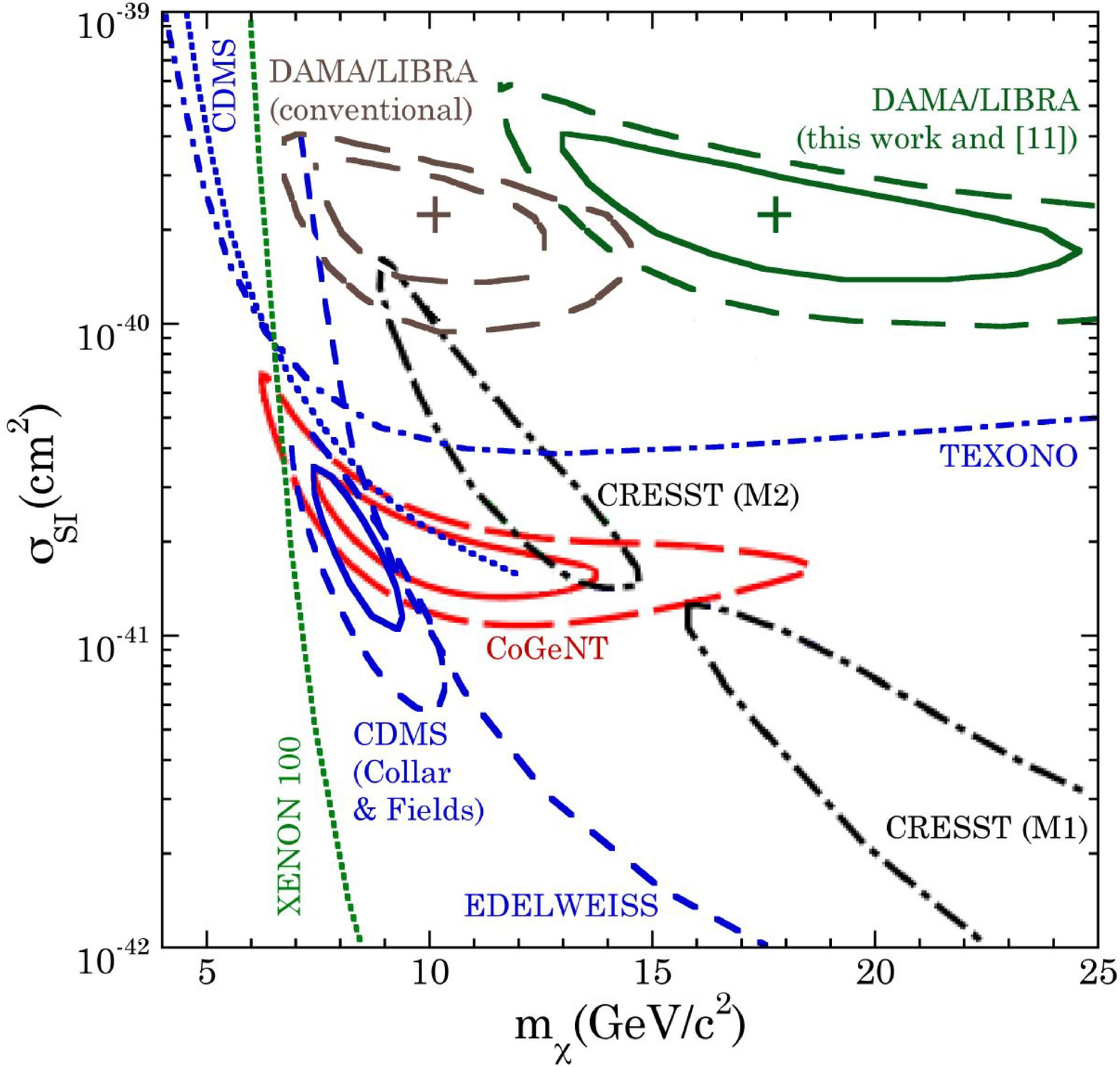}
\caption{\label{fig:measurements} DAMA/LIBRA ROI in spin-independent coupling vs.\ WIMP mass phase space, resulting from the adoption of the quenching factors derived from present measurements and those in \protect\cite{DD}, compared to the ROI generated by a conventional  $Q_{Na}\!=\!$ 0.3, $Q_{I}\!=\!$ 0.09 \protect\cite{damaquenching}. Solid contours correspond to a 90\% C.L. region, dashed to 99\% C.L. A Maxwellian galactic halo is assumed, with local parameters $v_{0}=$230 km/s,  $v_{esc}=$600 km/s, $\rho=$0.3 GeV/c$^{2}$cm$^{3}$. A CoGeNT ROI is from \protect\cite{nealdan2} and includes a correction for residual surface backgrounds as in \protect\cite{longcogent}. A possible ROI for CDMS is from \protect\cite{caf}.  ROIs and limits from CRESST \protect\cite{cresst}, CDMS \protect\cite{prevcdms}, EDELWEISS \protect\cite{edelweiss}, TEXONO \protect\cite{texono}, and XENON100 \protect\cite{lastxenon} are also shown.}
\end{figure}

A solid histogram in Fig.\ 2 represents the simulated nuclear recoil response following adoption of the NaI[Tl] quenching factors most recently measured in \cite{DD}, also introducing a simple model for the energy resolution based on the measured response to $^{57}$Co. The simulation includes self-shielding at the source and the (negligible) effect of known chemical impurities in the lead used. This prediction is found to be in good agreement with the data, even before any attempt at optimizing the energy dependence of the quenching factor or the resolution. A dashed histogram represents the prediction using the considerably larger quenching factors \cite{damaquenching} typically adopted in a WIMP interpretation of the DAMA/LIBRA results. It is at clear variance with present data. It should be noted that these conventional values, $Q_{\rm Na} = 0.30 \pm 0.01$ and $Q_{\rm I} = 0.09 \pm 0.01$ arise from an 
averaged response over 6.5 to 97 keVnr and  22 to 330 keVnr, respectively \cite{damaquenching}, i.e., from recoil energy regions much broader than those of actual interest for low-mass WIMP studies.

\begin{figure}[!htbp]
\includegraphics[width=0.45\textwidth]{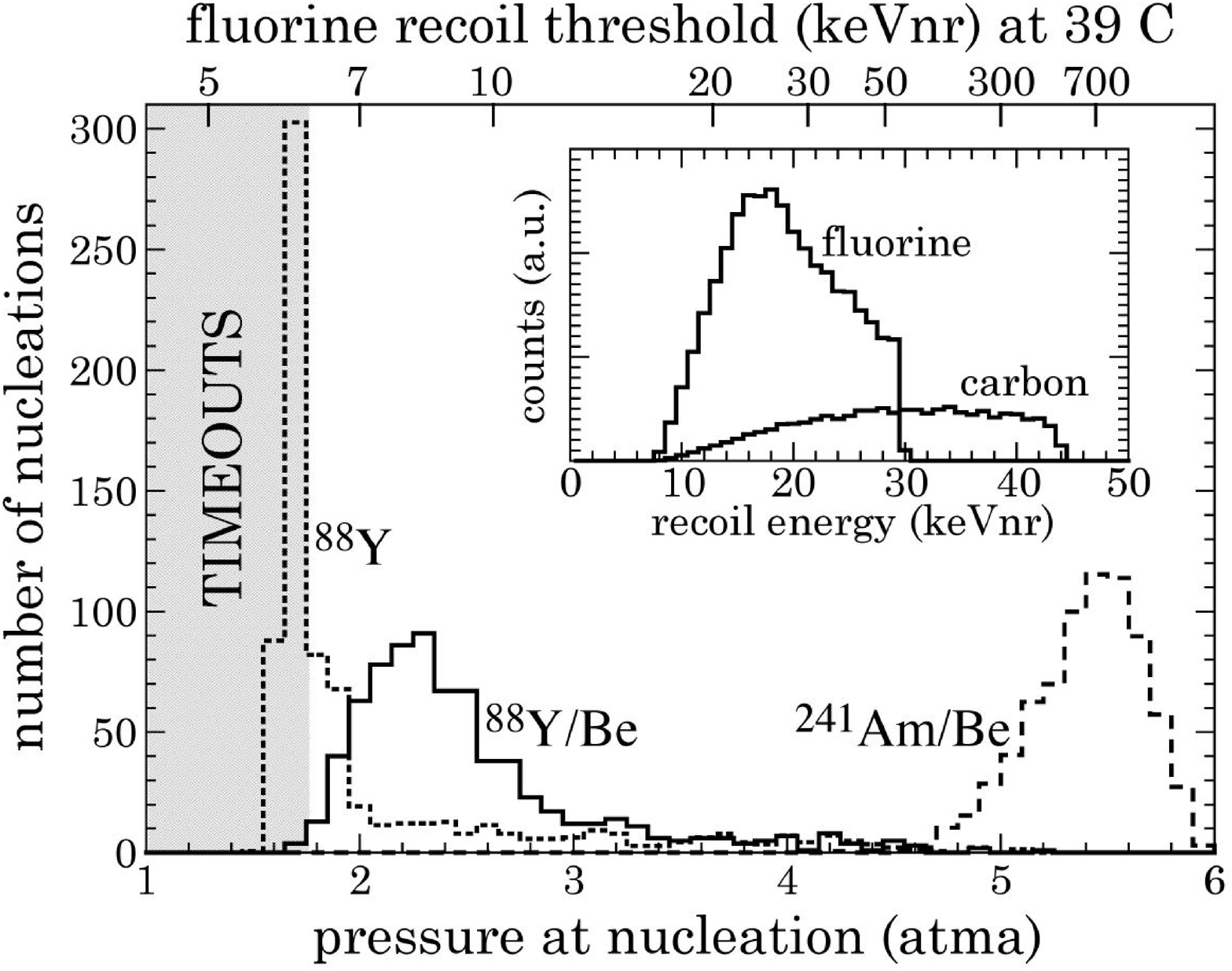}
\caption{\label{fig:measurements} Measured response of superheated CF$_{3}$I to $^{241}$Am/Be ($<$E$_{n}\!>=$ 4.5 MeV)  and $^{88}$Y/Be neutrons (E$_{n}=$ 0.152 MeV). As expected upon removal of the Be converter, only a minimal sensitivity to $^{88}$Y gammas is observed, starting at very high degrees of superheat \protect\cite{coupp,couppinprep}. Inset: simulated distribution of $^{88}$Y/Be nuclear recoil energies explored by the decompression protocol. These distributions can be skewed towards higher or lower mean energies by adjusting the decompression rate of the bubble chamber. No contribution from iodine recoils is expected in the slow decompression protocol employed.}
\end{figure}

Fig.\ 3 shows the sizable displacement in the DAMA/LIBRA region of interest (ROI) that stems from the adoption of  smaller quenching factors like those measured here and in \cite{DD}. The best-fit WIMP mass able to reproduce the DAMA/LIBRA annual modulation effect is shifted from $\sim$ 10 GeV/c$^{2}$ to $\sim$ 18 GeV/c$^{2}$ (crosses in the figure), pushing this ROI further into the region of parameter space excluded by numerous other experiments, and away from other anomalies. The reader is referred to \cite{DD} for an  additional discussion on astrophysical uncertainties perhaps capable of easing this now enhanced tension. Systematic effects affecting previous measurements of quenching factors at few keVnr, able to generate artificially large values, are described in \cite{DD,xecritique1}.

Fig.\ 4 displays results from the $^{88}$Y/Be irradiation of a small bubble chamber built by the author to study the response of moderately superheated CF$_{3}$I to low-energy nuclear recoils. In this particular case, no lead shielding is necessary, given the very large intrinsic insensitivity of this target to electron recoils \cite{coupp}. The protocol employed in these measurements is that of a slow controlled decompression of the chamber in the presence of the source, with or without Be converter, effectively scanning the theoretically-predicted nuclear recoil energy threshold. As this threshold is lowered, a "hot border" \cite{hot} is encountered when the maximum recoil energy produced by the monochromatic neutrons is reached, inducing bubble nucleations. The pressure cycle is then repeated. This protocol allows to extract the exact energy-dependence for this threshold from the analysis of the data \cite{couppinprep}.

Lastly, Fig.\ 5 shows the expected response to a lead-shielded $^{88}$Y/Be source in a small liquid xenon (LXe) calibration cell comparable to that in \cite{plante}. Overlapped is the very similar expected response to low-mass WIMPs like those of interest vis-\`a-vis a dark matter interpretation of the anomalies in \cite{DAMA,cogent,cresst}. The use of this source {\it in situ} with a larger LXe chamber such as the XENON100 dark matter detector would incorporate any instrumental effects (e.g., light collection efficiency, hardware trigger configuration, etc.) into the calibration of the detector response, but would also lead to higher-energy signals from multiple neutron scattering. These can however be identified through the good $\sim$1 mm spatial reconstruction of this detector \cite{rafi}. This straightforward calibration would provide a definitive test of the response of the XENON100 apparatus to low-mass WIMPs, resolving the controversy around its claimed sensitivity to WIMP masses below $\sim$ 12 GeV/c$^{2}$ \cite{gerbier2}.

\begin{figure}[!htbp]
\includegraphics[width=0.45\textwidth]{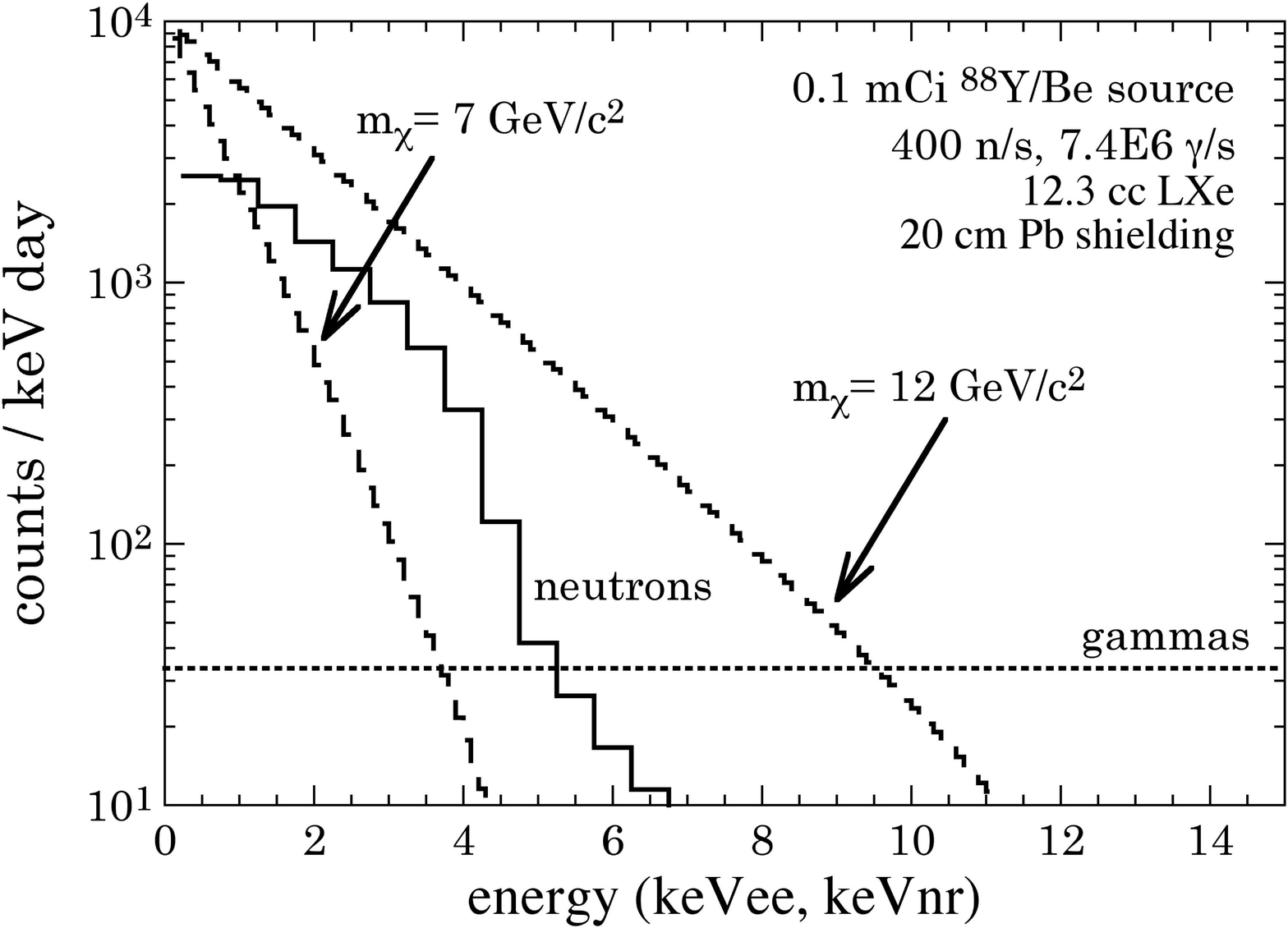}
\caption{\label{fig:measurements} Simulated response of a small LXe detector to an $^{88}$Y/Be source (nuclear recoil energies in keVnr, electron recoil energies from gamma component in keVee). A very similar spectral response to low-mass WIMPs like those of present interest is shown, normalized to an arbitrary cross-section.}
\end{figure}

In conclusion, a new and convenient calibration method able to replicate the expected signals from low-mass WIMP scattering has been demonstrated. Its use with a NaI[Tl] target confirms recent quenching factor measurements by the author using an independent approach \cite{DD}. These results substantially increase the difficulty in interpreting the DAMA/LIBRA anomaly within a WIMP dark matter context, under the standard particle physics and astrophysics phenomenological conventions. In an upcoming publication \cite{cosi}, $^{88}$Y/Be will be used to study the decay time of scintillation light from nuclear recoils in CsI[Na], for the few keVnr energy range expected from coherent neutrino scattering off nuclei, another possible realm of application.

The author thanks the Aspen Center for Physics for its hospitality during the completion of this Letter. The ACP is supported by the National Science Foundation under grant PHY-1066293.

\end{document}